%% file: sample-journal.tex
 \documentclass[format=acmsmall, review=false, screen=true]{acmart}

\usepackage{atbegshi}
\AtBeginDocument{\AtBeginShipoutNext{\AtBeginShipoutDiscard}}
\acmJournal{TMIS}
\acmVolume{9}
\acmNumber{0}
\acmArticle{0}
\acmYear{2018}
\acmMonth{0}

\setcopyright{acmcopyright}

\acmDOI{0000001.0000001}

\begin{document}
\title{Blockchains for Business Process Management - Challenges and Opportunities}

\author{Jan Mendling}
\orcid{0000-0002-7260-524X}
\affiliation{
  \institution{Wirtschaftsuniversit{\"a}t Wien}
  \city{Vienna}
  \country{Austria}
}
\email{jan.mendling@wu.ac.at}

\author{Ingo Weber}
\affiliation{
  \institution{Data61, CSIRO}
  \city{Sydney}
  \country{Australia}
}
\email{ingo.weber@data61.csiro.au}

\author{Wil van der Aalst}
\affiliation{
  \institution{Eindhoven University of Technology}
  \city{Eindhoven}
  \country{The Netherlands}
}
\email{w.m.p.v.d.aalst@tue.nl}

\author{Jan vom Brocke}
\affiliation{
  \institution{University of Liechtenstein}
  \city{Vaduz}
  \country{Liechtenstein}
}
\email{jan.vom.brocke@uni.li}

\author{Cristina Cabanillas}
\affiliation{
  \institution{Wirtschaftsuniversit{\"a}t Wien}
  \city{Vienna}
  \country{Austria}
}
\email{cristina.cabanillas@wu.ac.at}

\author{Florian Daniel}
\affiliation{
  \institution{Politecnico di Milano}
  \city{Milan}
  \country{Italy}
}
\email{florian.daniel@polimi.it}

\author{S\o ren Debois}
\affiliation{
  \institution{IT University of Copenhagen}
  \city{Copenhagen}
  \country{Denmark}
}
\email{debois@itu.dk}

\author{Claudio Di Ciccio}
\affiliation{
  \institution{Wirtschaftsuniversit{\"a}t Wien}
  \city{Vienna}
  \country{Austria}
}
\email{claudio.di.ciccio@wu.ac.at}

\author{Marlon Dumas}
\affiliation{
  \institution{University of Tartu}
  \city{Tartu}
  \country{Estonia}
}
\email{marlon.dumas@ut.ee}

\author{Schahram Dustdar}
\affiliation{
  \institution{TU Wien}
  \city{Vienna}
  \country{Austria}
}
\email{dustdar@tuwien.ac.at}

\author{Avigdor Gal}
\affiliation{
  \institution{Technion - Israel Institute of Technology}
  \city{Haifa}
  \country{Israel}
}
\email{avigal@technion.ac.il}

\author{Luciano Garc\'{\i}a-Ba\~nuelos}
\affiliation{
  \institution{University of Tartu}
  \city{Tartu}
  \country{Estonia}
}
\email{luciano.garcia@ut.ee}

\author{Guido Governatori}
\affiliation{
  \institution{Data61, CSIRO}
  \city{Brisbane}
  \country{Australia}
}
\email{guido.governatori@data61.csiro.au}

\author{Richard Hull}
\affiliation{
  \institution{IBM Research}
  \city{Yorktown Heights}
  \country{United States of America}
}
\email{hull@us.ibm.com}

\author{Marcello La Rosa}
\affiliation{
  \institution{Queensland University of Technology}
  \city{Brisbane}
  \country{Australia}
}
\email{m.larosa@qut.edu.au}

\author{Henrik Leopold}
\affiliation{
  \institution{Vrije Universiteit}
  \city{Amsterdam}
  \country{The Netherlands}
}
\email{h.leopold@vu.nl}

\author{Frank Leymann}
\affiliation{
  \institution{IAAS, Universit{\"a}t Stuttgart}
  \city{Stuttgart}
  \country{Germany}
}
\email{frank.leymann@iaas.uni-stuttgart.de}

\author{Jan Recker}
\affiliation{
  \institution{Queensland University of Technology}
  \city{Brisbane}
  \country{Australia}
}
\email{j.recker@qut.edu.au}

\author{Manfred Reichert}
\affiliation{
  \institution{Ulm University}
  \city{Ulm}
  \country{Germany}
}
\email{manfred.reichert@uni-ulm.de}

\author{Hajo A. Reijers}
\affiliation{
  \institution{Vrije Universiteit}
  \city{Amsterdam}
  \country{The Netherlands}
}
\email{h.a.reijers@vu.nl}

\author{Stefanie Rinderle-Ma}
\affiliation{
  \institution{University of Vienna}
  \city{Vienna}
  \country{Austria}
}
\email{stefanie.rinderle-ma@univie.ac.at}

\author{Andreas Solti}
\affiliation{
  \institution{Wirtschaftsuniversit{\"a}t Wien}
  \city{Vienna}
  \country{Austria}
}
\email{andreas.rogge-solti@wu.ac.at}

\author{Michael Rosemann}
\affiliation{
  \institution{Queensland University of Technology}
  \city{Brisbane}
  \country{Australia}
}
\email{m.rosemann@qut.edu.au}

\author{Stefan Schulte}
\affiliation{
  \institution{TU Wien}
  \city{Vienna}
  \country{Austria}
}
\email{stefan.schulte@tuwien.ac.at}

\author{Munindar P. Singh}
\affiliation{
  \institution{North Carolina State University}
  \city{Raleigh}
  \country{United States of America}
}
\email{mpsingh@ncsu.edu}

\author{Tijs Slaats}
\affiliation{
  \institution{University of Copenhagen}
  \city{Copenhagen}
  \country{Denmark}
}
\email{slaats@di.ku.dk}

\author{Mark Staples}
\affiliation{
  \institution{Data61, CSIRO}
  \city{Sydney}
  \country{Australia}
}
\email{mark.staples@data61.csiro.au}

\author{Barbara Weber}
\affiliation{
  \institution{Technical University of Denmark}
  \city{Lyngby}
  \country{Denmark}
}
\email{bweb@dtu.dk}

\author{Matthias Weidlich}
\affiliation{
  \institution{Humboldt-Universit{\"a}t zu Berlin}
  \city{Berlin}
  \country{Germany}
}
\email{matthias.weidlich@hu-berlin.de}

\author{Mathias Weske}
\affiliation{
  \institution{Hasso-Plattner-Institute, Universit{\"a}t Potsdam}
  \city{Potsdam}
  \country{Germany}
}
\email{mathias.weske@hpi.de}

\author{Xiwei Xu}
\affiliation{
  \institution{Data61, CSIRO}
  \city{Sydney}
  \country{Australia}
}
\email{xiwei.xu@data61.csiro.au}

\author{Liming Zhu}
\affiliation{
  \institution{Data61, CSIRO}
  \city{Sydney}
  \country{Australia}
}
\email{liming.zhu@data61.csiro.au}

\renewcommand\shortauthors{Mendling, J. et al}

\begin{abstract}
Blockchain technology offers a sizable promise to rethink the way inter-organizational business processes are managed because of its potential to realize execution without a central party serving as a single point of trust (and failure). To stimulate research on this promise and the limits thereof, in this paper we outline the challenges and opportunities of blockchain for Business Process Management (BPM). We first reflect how blockchains could be used in the context of the established BPM lifecycle and second how they might become relevant beyond. We conclude our discourse with a summary of seven research directions for investigating the application of blockchain technology in the context of BPM.
\end{abstract}

%
%


%
%


\maketitle

\input{intro}
\input{background}
\input{lifecycle}

\input{capab}

\section{Seven Future Research Directions}\label{Discu}
Blockchains will fundamentally shift how we deal with transactions in general, and therefore how organizations manage their business processes within their network. Our discussion of challenges in relation to the BPM lifecycle and beyond points to seven major future research directions. For some of them we expect viable insights to emerge sooner, for others later. The order loosely reflects how soon such insights might appear.
\begin{enumerate}
\item Developing a diverse set of \textit{execution and monitoring systems} on blockchain. Research in this area will have to demonstrate the feasibility of using blockchains for process-aware information systems. Among others, design science and algorithm engineering will be required here. Insights from software engineering and distributed systems will be informative.
\item Devising new \textit{methods for analysis and engineering} business processes based on blockchain technology. Research in this topic area will have to investigate how blockchain-based processes can be efficiently specified and deployed. Among others, formal research methods and design science will be required to study this topic. Insights from software engineering and database research will be informative here.
\item \textit{Redesigning processes} to leverage the opportunities granted by blockchain. Research in this context will have to investigate how blockchain may allow re-imagining specific processes and the collaboration with external stakeholders. The whole area of choreographies may be re-vitalized by this technology. Among others, design science will be required here. Insights from operations management and organizational science will be informative.
\item Defining appropriate methods for \textit{evolution and adaptation}. Research in this area will have to investigate the potential guarantees that can be made for certain types of evolution and adaptation. Among others, formal research methods will be required here. Insights from theoretical computer science and verification will be informative.
\item Developing techniques for identifying, discovering, and analyzing relevant processes for the \textit{adoption} of blockchain technology. Research on this topic will have to investigate which characteristics of blockchain as a technology best meet requirements of specific processes. Among others, empirical research methods and design science will be required. Insights from management science and innovation research will be informative here.
\item Understanding the \textit{impact on strategy and governance} of blockchains, in particular regarding new business and governance models enabled by revolutionary innovation based on blockchain. Research in this topic area will have to study which processes in an enterprise setting could be onoverganized differently using blockchain and which consequences this brings. Among others, empirical research methods will be required to investigate this topic. Insights from organizational science and business research will be informative.
\item Investigating the \textit{culture shift} towards openness in the management and execution of business processes, and on hiring as well as upskilling people as needed. Research in this topic area will have to investigate how corporate culture changes with the introduction of blockchains, and in how far this differs from the adoption of other technologies. Among others, empirical methods will be required for research in this area. Insights from organizational science and business research will be informative.
\end{enumerate}
The BPM and the Information Systems community have a unique opportunity to help shape this fundamental shift towards a distributed, trustworthy infrastructure to promote inter-organizational processes. With this paper we aim to provide clarity, focus, and impetus for the research challenges that are upon us.

\bibliographystyle{ACM-Reference-Format}
\bibliography{biblio}
\end{document}

%% file: intro.tex
\section{Introduction}
\label{sec:intro}

Business process management (BPM) is concerned with the design, execution, monitoring, and improvement of business processes. Systems that support the enactment and execution of processes 
have extensively been used by companies to streamline and automate \textit{intra}-organizational processes. Yet, for \textit{inter}-organizational processes, challenges of joint design and a lack of mutual trust have hampered a broader uptake.

Emerging \textit{blockchain} technology has the potential to drastically change the environment in which inter-organizational processes are able to operate. Blockchains offer a way to execute processes in a trustworthy manner even in a network without any mutual trust between nodes. Key aspects are specific algorithms that lead to consensus among the nodes and market mechanisms that motivate the nodes to progress the network. Through these capabilities, this technology has the potential to shift the discourse in BPM research about how systems might enable the enactment, execution, monitoring or improvement of business process within or across business networks.

In this paper, we describe what we believe are the main new challenges and opportunities of blockchain technology for BPM. This leads to directions for research activities to investigate both challenges and opportunities.
Section~\ref{sec:background} provides a background on fundamental concepts of blockchain technology and an illustrative example of how this technology applies to business processes.
Section~\ref{sec:bpm-lc} focuses on the impact of blockchains on the traditional \textit{BPM lifecycle phases}~\citep{DBLP:books/daglib/0031128}. Section~\ref{sec:capab} goes beyond it and asks which impact blockchains might have on core capability areas of BPM~\citep{rosemann2015six}.
Section~\ref{Discu} summarizes this discussion by emphasizing seven future research directions.

%% file: background.tex
\section{Background}
\label{sec:background}
This section summarizes the essential aspects of blockchain technology and discusses initial research efforts at the intersection of BPM and blockchains.
\subsection{Blockchain Technology}
In its original form, Blockchain is a distributed database technology that builds on a tamper-proof list of timestamped transaction records. Among others, it is used for cryptocurrencies such as Bitcoin~\citep{nakamoto2008bitcoin}.
Its innovative power stems from allowing parties to transact with others they do not trust over a computer network in which nobody is trusted. This is enabled by a combination of peer-to-peer networks, consensus-making, cryptography, and market mechanisms.

Blockchain derives its name from the fact that its essential data structure is a chained list of blocks. This chain of blocks is distributed over a peer-to-peer network, in which every node maintains the latest version of it. Blocks can contain information about transactions. In this way, we can for instance know that a buyer has ordered 200 items of a particular type of material from a vendor at a specific time. When a new block is added to the blockchain, it is signed using cryptographic methods. In this way, it can be checked if its content and its signature match. For example, if we take the content $c=$"Buyer orders 200 items from vendor" and apply a specific hash function $h(c)$, we get a unique result $r$. Every block is associated with a hash generated from its content \emph{and} the hash value of the previous block in the list. Hash values thus uniquely represent not only the transactions within blocks but also the ordering of every block. This mechanism is at the basis of the chain. In case somebody would try to alter a transaction, this would change the hash value of its block, and therefore break the chain. Since every node can create blocks in a peer-to-peer network, there has to be consensus on the new version of the blockchain including a new block.
This is achieved with consensus algorithms that are based on concepts like proof-of-work or proof-of-stake~\citep{Bentov2016}, and more recently \textit{proof-of-elapsed-time}\footnote{Intel: Proof of elapsed time (PoET). Available from  \url{http://intelledger.github.io/}}.
In proof-of-work, miners guess a value for a specific field, to fulfill the condition that  $r$ must be smaller than a threshold (which is dynamically adjusted by the network based on a predefined protocol). In proof-of-stake, miner selection considers the size of their stake , i.e., amount of cryptocurrency held by them. The rationale is that a high stake is a strong motivation for not cheating: if the miners cheat (and this is detected), the respective cryptocurrency will be devalued.
The network protocols and dynamic adjustment of thresholds are designed to avoid network overload.
In summary, these foundational blockchain concepts support two important notions that are also essential for business processes: the blockchain as a (tamper-proof) data structure captures the history and the current state of the network and transactions move the system to a new state.

Blockchain offers an additional concept that is important for business processes, called \textit{smart contracts}~\citep{szabo1997formalizing}. Consider again the example of the buyer ordering 200 items from the vendor. Business processes are subject to rules on how to respond to specific conditions. If, for instance, the vendor does not deliver within two weeks, the buyer might be entitled to receive a penalty payment. Such business rules can be expressed by smart contracts. For instance, the \textit{Ethereum} blockchain supports a Turing-complete programming language for smart contracts\footnote{\url{https://www.ethereum.org/}}. The code in these languages is deterministic and relies on a closed-world assumption: only information that is stored on the blockchain is available in the runtime environment. Smart contract code is deployed with a specific type of transaction. As with any other blockchain transaction, the deployment of smart contract code to the blockchain is immutable.
Once deployed, smart contracts offer a way to execute code directly on the blockchain network, like the conditional transfer of money in our example if a certain condition is fulfilled.

By using blockchain technology,
untrusted parties can establish trust in the truthful execution of the code. Smart contracts can be used to implement business collaborations in general and inter-organizational business processes in particular. The potential of blockchain-based distributed ledgers to enable collaboration in open environments has been successfully tested in diverse fields ranging from diamonds trading to securities settlement~\citep{walport2016distributed}.

At this stage, it has to be noted that blockchain technology still faces numerous general technological challenges. A mapping study by \cite{Yli-Huumo:2016:PlosOne} found that a majority of these challenges have not been addressed by the research community, albeit we note that blockchain developer communities actively discuss some of these challenges and suggest a myriad of potential solutions\footnote{\url{http://www.the-blockchain.com/2017/01/24/adi-ben-ari-outstanding-challenges-blockchain-technology-2017/}}.
Some of them can be addressed by using private or consortium blockchain instead of a fully open network~\citep{mougayar2016business}. In general, the technological challenges include the following~\cite{swan2015blockchain}.
\begin{description}
\item{\textbf{Throughput}} in the Ethereum blockchain is limited to approx.~15 transaction inclusions per second (tps) currently. In comparison, transaction volumes for the VISA payment network are 2,000 tps on average, with a tested capacity of up to 50,000 tps. However, the experimental Red Belly Blockchain which particularly caters to private or consortium blockchains has achieved more than 400,000 tps in a lab test\footnote{\url{http://poseidon.it.usyd.edu.au/~concurrentsystems/rbbc/}}.
\item{\textbf{Latency}} is also an issue. Transaction inclusion in the absence of network congestion takes a certain amount of time. In addition, a number of confirmation blocks are typically recommended to ensure the transaction does not get removed due to accidental or malicious forking. That means that transactions can be seen as committed after 60 minutes on average in Bitcoin, or 3 to 10 minutes in Ethereum. Even with improvements of techniques like the \emph{lightning network} or \emph{side chains} spawned off from the main chain, blockchains are unlikely to achieve latencies as low as centrally-controlled systems.
\item{\textbf{Size and bandwidth}} limitations are variations of the throughput issue: if the transaction volume of VISA were to be processed by Bitcoin, the full replication of the entire blockchain data structure would pose massive problems. \cite{Yli-Huumo:2016:PlosOne} quote 214 PB per year, thus posing a challenge in data storage and bandwidth. Private and consortium chains and concepts like the lightning network or side chains all aim to address these challenges. In this context it is worth noting that most everyday users can use \emph{wallets} instead, which require only small amounts of storage.
\item{\textbf{Usability}} is limited at this point, in terms of both developer support (lack of adequate tooling) and end-user support (hard to use and understand). Recent advances on developer support include efforts by some of the authors towards model-driven development of blockchain applications~\citep{Weber:2016:BPM,Garcia:2017:BPM,Tran:2017:CAISE}.
\item{\textbf{Security}} will always pose a challenge on an open network like a public blockchain. Security is often discussed in terms of the CIA properties~\citep{dhillon2000technical}. First, \textit{confidentiality} is per se low in a distributed system that replicates all data over its network, but can be addressed by targeted encryption~\citep{7546538}. Second, \textit{integrity} is a strong suit of blockchains, albeit challenges do exist~\citep{Eyal2014,Gervais:2016:SPP:2976749.2978341}. Third, \textit{availability} can be considered high in terms of reads from blockchain due to the wide replication, but is less favorable in terms of write availability~\cite{Weber:2017:SRDS}.
New attack vectors exist around forking, e.g., through network segregation~\cite{NG:2017:DSN}. These are particularly relevant in private or consortium blockchains.
\item{\textbf{Wasted resources,}} particularly electricity, are due to the consensus mechanism, where miners constantly compete in a race to mine the next block for a high reward. In an empirical analysis, \cite{Weber:2017:SRDS} found that about 10\% of announced new blocks on the Ethereum network were uncles (forks of length 1). This can be seen as wasteful, but is just a small indication of the vast duplication of effort in \emph{proof-of-work} mechanisms. Longer forks (at most of length 3) were extremely rare, so accidental forking seems unlikely in a well-connected network like the Internet -- but could occur if larger nations were cut off temporarily or even permanently. Alternatives to the proof-of-work, like \emph{proof-of-stake}~\citep{Bentov2016}, have been discussed for a while and would be much more efficient. At the time of writing, they remain an unproven but highly interesting alternative. Proof-of-work makes very low assumptions in trusting other participants, which is well suited for an open network managing digital assets. Designing more efficient protocols without relaxing these assumptions has proven a challenge.
\item{\textbf{Hard forks}} are changes to the protocol of a blockchain which enable transactions or blocks which were previously considered invalid~\citep{conf/p2p/DeckerW13}. They essentially change the rules of the game and therefore require adoption by a vast majority of the miners to be effective~\citep{7163021}. While hard forks can be controversial in public blockchains, as demonstrated by the split of the Ethereum blockchain into a hard forked main chain and Ethereum Classic (ETC), this is less of an issue for private and consortium blockchains where such a consensus is more easily found.
\end{description}
Many of these general technological challenges of blockchains are currently the focus of the emerging body of research. As noted, our main interest is in the \textit{potential} of blockchain technology to enable a shift in BPM research. Our belief is vested both in the novel technological properties discussed above and in the already available attempts of using blockchain technology in the definition and implementation of fundamentally novel business processes. We review these attempts in the following.

\subsection{Business Processes and Blockchain Technology}
We are not the first to identify the application potential of blockchain technology to business processes. In fact, several blockchains are currently adopted in various domains to facilitate the operation of new business processes. For example, \cite{DBLP:journals/bise/NoferGHS17} list applications in the financial sector including cryptocurrency transactions, securities trading and settlement, and insurances as well as non-financial applications such as notary services, music distribution, and various services like proof of existence, authenticity, or storage. Other works describe application scenarios involving blockchain technology in logistics and supply chain processes, for instance in the agricultural sector~\citep{Risks-Blockchain-2017}.

A proposal to support inter-organizational processes through blockchain technology is described by~\cite{Weber:2016:BPM}: large parts of the control flow and business logic of inter-organizational business processes can be compiled from process models into smart contracts which ensure the joint process is correctly executed. So-called \textit{trigger} components allow connecting these inter-organizational process implementations to Web services and internal process implementations. These triggers serve as a bridge between the blockchain and enterprise applications.
The cryptocurrency concept enables the optional implementation of conditional payment and built-in escrow management at defined points within the process, where this is desired and feasible.

\begin{figure}[t]
\begin{center}
\includegraphics[width=1\textwidth]{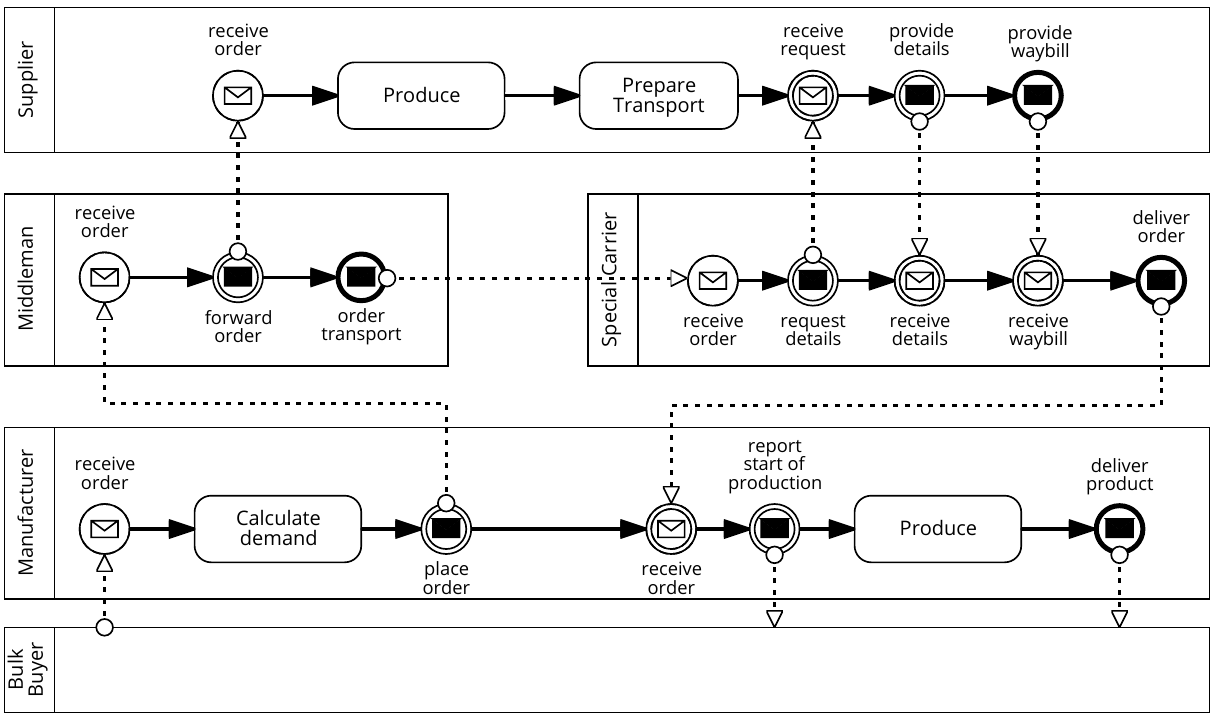}
\caption{Supply Chain Scenario from~\cite{Weber:2016:BPM}}\label{fig:collaboration}
\end{center}
\end{figure}

\begin{figure}[t]
\begin{center}
\includegraphics[width=.8\textwidth]{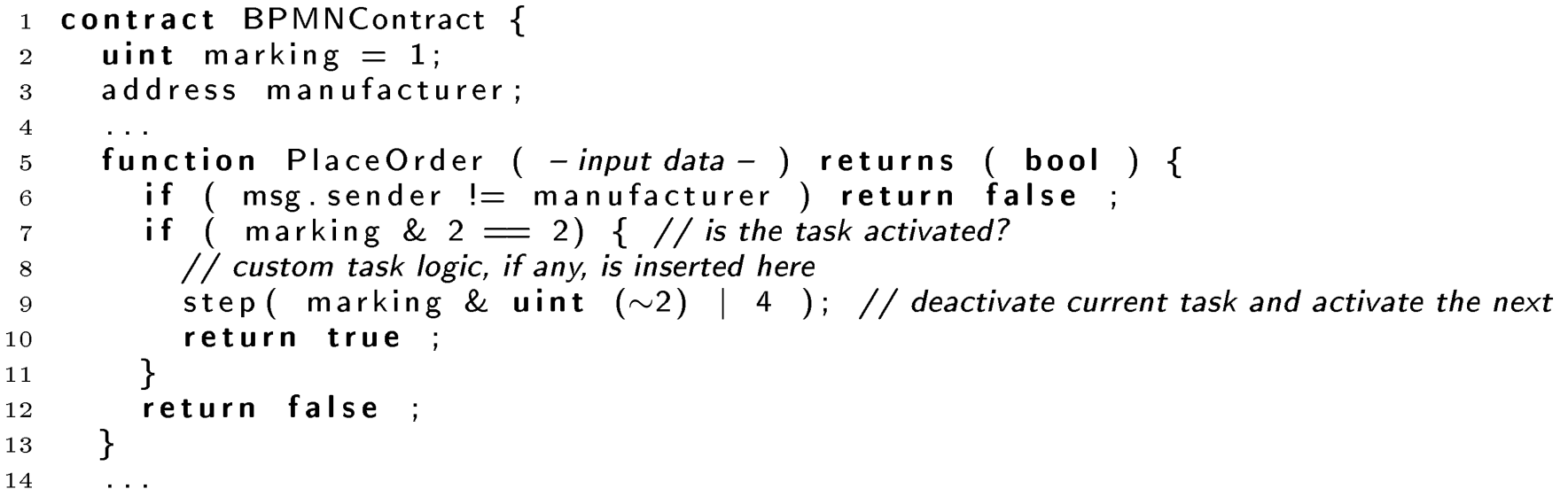}
\caption{Smart contract snippet illustrating how code is generated from a BPMN model. It shows the implementation of function \texttt{PlaceOrder} from the above process model. This function is to be executed by the Manufacturer, which is checked in line 6. Subsequently, we check if the function is activated in line 7. If so, any custom task logic is executed, and the activation of tasks is updated in line 9. For more details, see~\cite{Garcia:2017:BPM}.}\label{fig:code}
\end{center}
\end{figure}

To illustrate these capabilities, Figure~\ref{fig:collaboration} shows a simplified supply chain scenario, where a bulk buyer orders goods from a manufacturer. The manufacturer, in turn, orders supplies through a middleman, which are sent from the supplier to the manufacturer via a special carrier.
Without global monitoring each participant has restricted visibility of the overall progress. This may very well be a basis for misunderstandings and shifting blame in cases of conflict. Model-driven approaches such as proposed by~\cite{Weber:2016:BPM,Garcia:2017:BPM} produce code of smart contracts that implement the process (see Figure~\ref{fig:code}).

If executed using smart contracts on a blockchain, typical barriers complicating the deployment of inter-organizational processes can be removed. 
(i) The blockchain can serve as an immutable public ledger, so that participants can review a trustworthy history of messages to pinpoint the source of an error. This means that all state-changing messages have to be recorded in the blockchain. (ii) Smart contracts can offer independent process monitoring from a global viewpoint, such that only expected messages are accepted, and only if they are sent from the player registered for the respective role in the process instance.
(iii) Encryption can ensure that only the data that must be visible is public, while the remaining data is only readable for the process participants that require it.

These capabilities demonstrate how blockchains can help organizations to implement and execute business processes across organizational boundaries even if they cannot agree on a trusted third party. 
This is a fundamental advance, because the core aspects of this technology enable support of enterprise collaborations going far beyond asset management, including the management of entire supply chains, tracking food from source to consumption to increase safety, or sharing personal health records in privacy-ensuring ways amongst medical service providers.

The technical realization of this advance is still nascent at this stage, although some early efforts can be found in the literature. For example, smart contracts that enforce a process execution in a trustworthy way can be generated from BPMN process models~\citep{Weber:2016:BPM} and from domain-specific languages~\citep{Frantz:ECAS:2016}. Further cost optimizations are proposed by~\cite{Garcia:2017:BPM}. Figure~\ref{fig:code} shows a code excerpt that was generated by this approach.
In a closely related work, \cite{DBLP:conf/icsoc/HullBCDHV16} emphasize the affinity of artifact-centric process specification~\citep{DBLP:journals/debu/CohnH09,DBLP:conf/bpm/MarinHV12} for blockchain execution.

Even at this stage, research on the benefits and potentials of blockchain technology is mixed with studies that highlight or examine issues and challenges. For example, \cite{Norta:2015:BIR,Norta:2016:ICACDS} discusses ways to ensure secure negotiation and creation of smart contracts for Decentralized Autonomous Organizations (DAOs), among others in order to avoid attacks like the DAO hack during which approx. US\$ 60M were stolen. This in turn was remediated 
by a hard fork of the Ethereum blockchain, which was controversial among the respective mining node operators and resulted in a part of the public Ethereum network splintering off into the \emph{Ethereum Classic} (ETC) network. This split, in turn, caused major issues for the network in the medium term, allowing among others \emph{replay attacks} where transactions from Ethereum can be replayed on ETC.
A formal analysis of smart contract participants using game theory and formal methods is conducted by~\cite{Bigi:2015:Degano}. As pointed out by~\cite{Norta:2016:ICACDS}, the assumption of perfect rationality underlying the game-theoretic analysis is unlikely to hold for human participants.

These examples show that blockchain technology and its application to BPM are at an important crossroads: technical realization issues blend with promising application scenarios; early implementations mix with unanticipated challenges. It is timely, therefore, to discuss in broad and encompassing ways where open questions lie that the scholarly community should be interested in addressing. We do so in the two sections that follow.

%% file: lifecycle.tex
\section{Blockchain Technology and the BPM Lifecycle}\label{sec:bpm-lc}
In this section, we discuss blockchain in relation to the traditional BPM lifecycle~\citep{DBLP:books/daglib/0031128} including the following phases:
identification, discovery, analysis, redesign, implementation, execution, monitoring, and adaptation.
Using the traditional BPM lifecycle as a framework of reference allows us to discuss many incremental changes that blockchains might provide.

\subsection{Identification}
Process identification is concerned with the high-level description and evaluation of a company from a process-oriented perspective, thus connecting strategic alignment with process improvement. Currently, identification is mostly approached from an inward-looking perspective~\citep{DBLP:books/daglib/0031128}.
Blockchain technology adds another relevant perspective for evaluating high-level processes in terms of the implied strengths, weaknesses, opportunities, and threats. For example, how can a company systematically identify the most suitable processes for blockchains or the most threatened ones?
Research is needed into how this perspective can be integrated into the identification phase. Because blockchains have affinity with the support of inter-organizational processes, process identification may need to encompass not only the needs of one organization, but broader known and even unknown partners.

\subsection{Discovery}
Process discovery refers to the collection of information about the current way a process operates and its representation as an \textit{as-is} process model. Currently, methods for process discovery are largely based on interviews, walkthroughs and documentation analysis, complemented with automated process discovery techniques over non-encrypted event logs generated by process-aware information systems~\citep{DBLP:books/sp/Aalst16}.
Blockchain technology defines new challenges for process discovery techniques: the information may be fragmented and encrypted; accounts and keys can change frequently; and payload data may be stored partly on-chain and partly off-chain.
For example, how can a company discover an overall process from blockchain transactions when these might not be logically related to a process identifier?
This fragmentation might require a repeated alignment of information from all relevant parties operating on the blockchain. Work on matching could represent a promising starting point to solve this problem~\citep{DBLP:books/daglib/0032976,gal2011uncertain,DBLP:conf/bpm/CayogluDDFGHKLLLMOSSTUWW13}. There is both the risk and opportunity of conducting process mining on blockchain data. An opportunity could involve  establishing trust in how a process or a prospective business partner operates, while a risk is that other parties might be able to understand operational characteristics from blockchain transactions.
There are also opportunities for reverse engineering business processes, among others, from smart contracts.

\subsection{Analysis}
Process analysis refers to obtaining insights into issues relating to the way
a business process currently operates. Currently, the analysis of processes mostly builds on data that is available inside of organizations or from perceptions shared by internal and external process stakeholders~\citep{DBLP:books/daglib/0031128}.
Records of processes executed on the blockchain yield valuable information that can help to assess the case load, durations, frequencies of paths, parties involved, and correlations between unencrypted data items. These pieces of information can be used to discover processes, detect deviations, and conduct root cause analysis~\citep{DBLP:books/sp/Aalst16}, 
ranging from small groups of companies to an entire industry at large.
The question is which effort is required to bring the available blockchain transaction data into a format that permits such analysis.

\subsection{Redesign}
Process redesign deals with the systematic improvement of a process. Currently, approaches like redesign heuristics build on the assumption that there are recurring patterns of how a process can be improved~\citep{DBLP:journals/bise/VanwerschSVVGPM16}. Blockchain technology offers novel ways of improving specific business processes or resolving specific problems.
For instance, instead of involving a trustee to release a payment if an agreed condition is met, a buyer and a seller of a house might agree on a smart contract instead. The question is where blockchains can be applied for optimizing existing interactions and where new interaction patterns without a trusted central party can be established, potentially drawing on insights from related research on Web service interaction~\citep{barros2005service}.
A promising direction for developing blockchain-appropriate abstractions and heuristics may come from data-aware workflows~\citep{DBLP:conf/bpm/MarinHV12} and BPMN choreography diagrams~\citep{DBLP:journals/is/DeckerW11}. Both techniques combine two primary ingredients of blockchain, namely data and process, in a holistic manner that is well-suited for top-down design of cross-organizational processes.
It might also be beneficial to formulate blockchain-specific redesign heuristics that could mimic how
Incoterms~\citep{ramberg2011icc} define standardized interactions in international trade. Specific challenges for redesign include the joint engineering of blockchain processes between all parties involved, an ongoing problem for choreography design.

\subsection{Implementation}
Process implementation refers to the procedure of transforming a \textit{to-be} model into software components executing the business process. Currently, business processes are often implemented using process-aware information systems or business process management systems inside single organizations.
In this context, the question is how can the involved parties make sure that the implementation that they deploy on the blockchain supports their process as desired.
Some of the challenges regarding the transformation of a process model to blockchain artifacts are discussed by~\cite{Weber:2016:BPM}. Several ideas from earlier work on choreography can be reused in this new setting~\citep{DBLP:conf/caise/AalstW01,mendling2008ws,2008-Weber-IJBPIM,DBLP:journals/is/DeckerW11,RE-14:Protos,AAMAS-Comma-12}.
It has to be noted that choreographies have not been adopted by industry to a large extent yet. Despite this, they are especially helpful in inter-organizational settings, where it is not possible to control and monitor a complete process in a centralized fashion because of organizational borders~\citep{breu13}. To verify that contracts between choreography stakeholders have been fulfilled, a trust basis, which is not under control of a particular party, needs to be established. Blockchains may serve to establish this kind of trust between stakeholders.

An important engineering challenge on the implementation level is the identification and definition of abstractions for the design of blockchain-based business process execution.
Libraries and operations for engines are required, accompanied by modeling primitives and language extensions of BPMN. Software patterns and anti-patterns will be of good help to engineers designing blockchain-based processes.
There is also a need for new approaches for quality assurance, correctness, and verification, as well as
for new corresponding correctness criteria.
These can build on existing notions of compliance~\citep{DBLP:journals/toit/AalstDORV08}, reliability~\cite{DBLP:conf/saint/SattanathanTNMM08}, quality of services~\citep{DBLP:journals/tse/ZengBNDKC04} or data-aware workflow verification~\citep{DBLP:conf/pods/CalvaneseGM13}, but will have to go further in terms of consistency and consideration of potential payments. Furthermore, dynamic partner binding and rebinding is a challenge that requires attention. Process participants will have to find partners, either manually or automatically on dedicated marketplaces using  dedicated look-up services. The property of inhabiting a certain role in a process might itself be a tradable asset. For example, a supplier might auction off the role of shipper to the highest bidder as part of the process. Finally, as more and more companies use blockchain, there will be a proliferation of smart contract templates available for use. Tools for finding templates appropriate for a given style of collaboration will be essential. All these characteristics emphasize the need for specific testing and verification approaches.

\subsection{Execution}
Execution refers to the instantiation of individual cases and their information-technological processing. Currently, such execution is facilitated by process-aware information systems or business process management systems~\citep{DBLP:books/daglib/0031128}.
For the actual execution of a process deployed on a blockchain following the method of~\cite{Weber:2016:BPM}, several differences with the traditional ways exist. 
During the execution of an instance, messages between participants need to be passed as blockchain transactions to the smart contract; resulting messages need to be observed from the blocks in the blockchain. Both of these can be achieved by integrating blockchain technology directly with existing enterprise systems or through the use of dedicated integration components, such as the triggers suggested by~\cite{Weber:2016:BPM}. First prototypes like Caterpillar as a BPMS that build on blockchains are emerging~\cite{DBLP:conf/bpm/Lopez-PintadoGD17}.
The main challenge here involves ensuring correctness and security, especially when monetary assets are transferred using this technology.

\subsection{Monitoring}
Process monitoring refers to collecting events of process executions, displaying them in an understandable way, and triggering alerts and escalation in cases where undesired behavior is observed.
Currently, such process execution data is recorded by systems that support process execution~\citep{DBLP:books/daglib/0031128}.
First, we face issues in terms of data fragmentation and encryption as in the analysis phase. For example, the data on the blockchain alone will likely not be enough to monitor the process, but require an integration with local off-chain data.
Once such tracing in place, the global view of the process can be monitored independently by each involved party.
This provides a suitable basis for continuous conformance and compliance checking and monitoring of service-level agreements.
Second, based on monitoring data exchanged via the blockchain, it is possible to verify if a process instance meets the original process model and the contractual obligations of all involved process stakeholders. For this, blockchain technology can be exploited to store the process execution data and handoffs between process participants. Notably, this is even possible without the usage of smart contracts, i.e., in a first-generation blockchain like the one operated by Bitcoin~\citep{PSHW17}.

\subsection{Adaptation and Evolution}
Runtime adaptation refers to the concept of changing the process during execution. In traditional approaches, this can for instance be achieved by allowing participants in a process to change the model during its execution~\citep{DBLP:books/daglib/0030179}.
Interacting partners might take a defensive stance in order to avoid certain types of adaptation.
As discussed by~\cite{Weber:2016:BPM}, blockchain can be used to enforce conformance with the model, so that participants can rely on the joint model being followed.
In such a setting, adaptation is by default something to be \emph{avoided:} if a participant can change the model, this could be used to gain an unfair advantage over the other participants. For instance, the rules of retrieving cryptocurrency from an escrow account could be changed or the terms of payment.
In this setting, process adaptation must strictly adhere to defined paths for it, e.g., any change to a deployed smart contract may require a transaction signed by all participants.
In contrast, the method proposed by~\citep{PSHW17} allows runtime adaptation, but assumes that relevant participants monitor the execution and react if a change is undesired.

If smart contracts enforce the process, there are also problems arising in relation to evolution: new smart contracts need to be deployed to reflect changes to a new version of the process model. Porting running instances from an old version to a new one would require effective coordination mechanisms involving all participants. Some challenges for choreographies are summarized by~\cite{DBLP:journals/is/FdhilaIRR15}.

%% file: capab.tex
\section{Blockchain Technology and BPM Capabilities}
\label{sec:capab}
There are also challenges and opportunities for BPM and blockchain technology beyond the classical BPM lifecycle. We refer to the BPM capability areas~\citep{rosemann2015six} beyond the methodological support we reflected above, including strategy, governance, information technology, people, and culture.

\subsection{Strategy}
Strategic alignment refers to the active management of connections between organizational priorities and business processes~\citep{rosemann2015six}, which aims at facilitating effective actions to improve business performance.
Currently, various approaches to BPM assume that the corporate strategy is defined first and business processes are aligned with the respective strategic imperatives~\citep{DBLP:books/daglib/0031128}.
Blockchain technology challenges these approaches to strategic alignment.
For many companies, blockchains define a potential threat to their core business processes. For instance, the banking industry could see a major disintermediation based on blockchain-based payment services~\citep{guo2016blockchain}. Also lock-in effects~\citep{tassey2000standardization} might deteriorate when, for example, the banking service is not the banking network itself anymore, but only the interface to it. These developments could lead to business processes and business models being under strong influence of technological innovations outside of companies.

\subsection{Governance}
BPM governance refers to appropriate and transparent accountability in terms of roles, responsibilities, and decision processes for different BPM-related programs, projects, and operations~\citep{rosemann2015six}.
Currently, BPM as a management approach builds on the explicit definition of BPM-related roles and responsibilities with a focus on the internal operations of a company.
Blockchain technology might change governance towards a more externally oriented model of self-governance based on smart contracts.
Research on corporate governance investigates agency problems and mechanisms to provide effective incentives for intended behavior~\citep{shleifer1997survey}. Smart contracts can be used to establish new governance models as exemplified by The Decentralized Autonomous Organization (The DAO)~\footnote{\url{https://daohub.org}}. It is an important question in how far this idea of The DAO can be extended towards reducing the agency problem of management discretion or eventually eliminate the need for management altogether.
Furthermore, the revolutionary change suggested by The DAO for organization shows just how disruptive this technology can be, and whether similarly radical changes could apply to BPM.

\subsection{Information Technology}
BPM-related information technology subsumes all systems that support process execution, such as process-aware information systems and business process management systems. These systems typically assume central control over the process.

Blockchain technology enables novel ways of process execution, but several challenges in terms of security and privacy have to be considered. While the visibility of encrypted data on a blockchain is restricted, it is up to the participants in the process to ensure that these mechanisms are used according to their confidentiality requirements. Some of these requirements are currently being investigated in the financial industry\footnote{\url{https://gendal.me/2016/04/05/introducing-r3-corda-a-distributed-ledger-designed-for-financial-services/}}. Further challenges can be expected with the introduction of the General Data Protection Regulation\footnote{\url{http://eur-lex.europa.eu/legal-content/EN/TXT/?uri=uriserv:OJ.L\_{}.2016.119.01.0001.01.ENG}}.
It is also not clear, which new attack scenarios on blockchain networks might emerge~\citep{hurlburt2016might}. Therefore, guidelines for using private, public, or consortium-based blockchains are required~\citep{mougayar2016business}. It also has to be decided what types of smart contract and which cryptocurrency are allowed to be used in a corporate setting.

\subsection{People}
People in this context refers to all individuals, possibly in different roles, who engage with BPM~\citep{rosemann2015six}.
Currently, these are people who work as process analyst, process manager, process owner or in other process-related roles. The roles of these individuals are shaped by skills in the area of management, business analysis and requirements engineering.
In this capability area, the use of blockchain technology requires extensions of their skill sets.
New required skills relate to partner and contract management, software enginering, and cryptography.
Also, people have to be willing to design blockchain-based collaborations within the frame of existing regulations to enable adoption. This implies that research into blockchain-specific technology acceptance is needed, extending the established technology acceptance model~\citep{venkatesh2003user}.

\subsection{Culture}
Organizational culture is defined by the collective values of a group of people in an organization~\citep{rosemann2015six}.
Currently, BPM is discussed in relation to organizational culture~\citep{brocke2011culture} from a perspective that emphasizes an affinity with clan and hierarchy culture~\citep{stemberger-2017}. These cultural types are often found in the many companies that use BPM as an approach for documentation. Blockchains are likely to influence organizational culture towards a stronger emphasis on flexibility and an outward-looking perspective. In the competing values framework by~\cite{cameron2005diagnosing}, these aspects are associated with an adhocracy organizational culture. Furthermore, not only consequences of blockchain adoption have to be studied, but also antecedants. These include organizational factors that facilitate early and successful adoption. 

%% file: sample-journal.bbl

\begin{thebibliography}{59}


\ifx \showCODEN    \undefined \def \showCODEN     #1{\unskip}     \fi
\ifx \showDOI      \undefined \def \showDOI       #1{#1}\fi
\ifx \showISBNx    \undefined \def \showISBNx     #1{\unskip}     \fi
\ifx \showISBNxiii \undefined \def \showISBNxiii  #1{\unskip}     \fi
\ifx \showISSN     \undefined \def \showISSN      #1{\unskip}     \fi
\ifx \showLCCN     \undefined \def \showLCCN      #1{\unskip}     \fi
\ifx \shownote     \undefined \def \shownote      #1{#1}          \fi
\ifx \showarticletitle \undefined \def \showarticletitle #1{#1}   \fi
\ifx \showURL      \undefined \def \showURL       {\relax}        \fi
\providecommand\bibfield[2]{#2}
\providecommand\bibinfo[2]{#2}
\providecommand\natexlab[1]{#1}
\providecommand\showeprint[2][]{arXiv:#2}

\bibitem[\protect\citeauthoryear{Barros, Dumas, and ter Hofstede}{Barros
  et~al\mbox{.}}{2005}]%
        {barros2005service}
\bibfield{author}{\bibinfo{person}{Alistair Barros}, \bibinfo{person}{Marlon
  Dumas}, {and} \bibinfo{person}{Arthur~HM ter Hofstede}.}
  \bibinfo{year}{2005}\natexlab{}.
\newblock \showarticletitle{Service interaction patterns}. In
  \bibinfo{booktitle}{{\em International Conference on Business Process
  Management}}. Springer, \bibinfo{pages}{302--318}.
\newblock


\bibitem[\protect\citeauthoryear{Bentov, Gabizon, and Mizrahi}{Bentov
  et~al\mbox{.}}{2016}]%
        {Bentov2016}
\bibfield{author}{\bibinfo{person}{Iddo Bentov}, \bibinfo{person}{Ariel
  Gabizon}, {and} \bibinfo{person}{Alex Mizrahi}.}
  \bibinfo{year}{2016}\natexlab{}.
\newblock \bibinfo{booktitle}{{\em Cryptocurrencies Without Proof of Work}}.
\newblock \bibinfo{publisher}{Springer Berlin Heidelberg},
  \bibinfo{address}{Berlin, Heidelberg}, \bibinfo{pages}{142--157}.
\newblock
\showISBNx{978-3-662-53357-4}
\showDOI{%
\url{https://doi.org/10.1007/978-3-662-53357-4_10}}


\bibitem[\protect\citeauthoryear{Bigi, Bracciali, Meacci, and Tuosto}{Bigi
  et~al\mbox{.}}{2015}]%
        {Bigi:2015:Degano}
\bibfield{author}{\bibinfo{person}{Giancarlo Bigi}, \bibinfo{person}{Andrea
  Bracciali}, \bibinfo{person}{Giovanni Meacci}, {and} \bibinfo{person}{Emilio
  Tuosto}.} \bibinfo{year}{2015}\natexlab{}.
\newblock \bibinfo{booktitle}{{\em Validation of Decentralised Smart Contracts
  Through Game Theory and Formal Methods}}.
\newblock \bibinfo{publisher}{Springer International Publishing},
  \bibinfo{pages}{142--161}.
\newblock


\bibitem[\protect\citeauthoryear{Bonneau, Miller, Clark, Narayanan, Kroll, and
  Felten}{Bonneau et~al\mbox{.}}{2015}]%
        {7163021}
\bibfield{author}{\bibinfo{person}{J. Bonneau}, \bibinfo{person}{A. Miller},
  \bibinfo{person}{J. Clark}, \bibinfo{person}{A. Narayanan},
  \bibinfo{person}{J.~A. Kroll}, {and} \bibinfo{person}{E.~W. Felten}.}
  \bibinfo{year}{2015}\natexlab{}.
\newblock \showarticletitle{SoK: Research Perspectives and Challenges for
  Bitcoin and Cryptocurrencies}. In \bibinfo{booktitle}{{\em 2015 IEEE
  Symposium on Security and Privacy}}. \bibinfo{pages}{104--121}.
\newblock
\showISSN{1081-6011}
\showDOI{%
\url{https://doi.org/10.1109/SP.2015.14}}


\bibitem[\protect\citeauthoryear{Breu, Dustdar, Eder, Huemer, Kappel,
  K\"{o}pke, Langer, Mangler, Mendling, Neumann, Rinderle-Ma, Schulte,
  Sobernig, and Weber}{Breu et~al\mbox{.}}{2013}]%
        {breu13}
\bibfield{author}{\bibinfo{person}{Ruth Breu}, \bibinfo{person}{Schahram
  Dustdar}, \bibinfo{person}{Johann Eder}, \bibinfo{person}{Christian Huemer},
  \bibinfo{person}{Gerti Kappel}, \bibinfo{person}{Julius K\"{o}pke},
  \bibinfo{person}{Philip Langer}, \bibinfo{person}{J\"{u}rgen Mangler},
  \bibinfo{person}{Jan Mendling}, \bibinfo{person}{Gustaf Neumann},
  \bibinfo{person}{Stefanie Rinderle-Ma}, \bibinfo{person}{Stefan Schulte},
  \bibinfo{person}{Stefan Sobernig}, {and} \bibinfo{person}{Barbara Weber}.}
  \bibinfo{year}{2013}\natexlab{}.
\newblock \showarticletitle{{Towards Living Inter-Organizational Processes}}.
  In \bibinfo{booktitle}{{\em 15th IEEE Conference on Business Informatics}}.
  \bibinfo{publisher}{IEEE}, \bibinfo{pages}{363--366}.
\newblock
\showDOI{%
\url{https://doi.org/dx.doi.org/10.1109/CBI.2013.59}}


\bibitem[\protect\citeauthoryear{Calvanese, {De Giacomo}, and
  Montali}{Calvanese et~al\mbox{.}}{2013}]%
        {DBLP:conf/pods/CalvaneseGM13}
\bibfield{author}{\bibinfo{person}{Diego Calvanese}, \bibinfo{person}{Giuseppe
  {De Giacomo}}, {and} \bibinfo{person}{Marco Montali}.}
  \bibinfo{year}{2013}\natexlab{}.
\newblock \showarticletitle{Foundations of data-aware process analysis: a
  database theory perspective}. In \bibinfo{booktitle}{{\em Proceedings of the
  32nd {ACM} {SIGMOD-SIGACT-SIGART} Symposium on Principles of Database
  Systems, {PODS} 2013, New York, NY, {USA} - June 22 - 27, 2013}},
  \bibfield{editor}{\bibinfo{person}{Richard Hull} {and}
  \bibinfo{person}{Wenfei Fan}} (Eds.). \bibinfo{publisher}{{ACM}},
  \bibinfo{pages}{1--12}.
\newblock


\bibitem[\protect\citeauthoryear{Cameron and Quinn}{Cameron and Quinn}{2005}]%
        {cameron2005diagnosing}
\bibfield{author}{\bibinfo{person}{Kim~S Cameron} {and}
  \bibinfo{person}{Robert~E Quinn}.} \bibinfo{year}{2005}\natexlab{}.
\newblock \bibinfo{booktitle}{{\em Diagnosing and changing organizational
  culture: Based on the competing values framework}}.
\newblock \bibinfo{publisher}{John Wiley \& Sons}.
\newblock


\bibitem[\protect\citeauthoryear{Cayoglu, Dijkman, Dumas, Fettke,
  Garc{\'{\i}}a{-}Ba{\~{n}}uelos, Hake, Klinkm{\"{u}}ller, Leopold, Ludwig,
  Loos, Mendling, Oberweis, Schoknecht, Sheetrit, Thaler, Ullrich, Weber, and
  Weidlich}{Cayoglu et~al\mbox{.}}{2014}]%
        {DBLP:conf/bpm/CayogluDDFGHKLLLMOSSTUWW13}
\bibfield{author}{\bibinfo{person}{Ugur Cayoglu}, \bibinfo{person}{Remco~M.
  Dijkman}, \bibinfo{person}{Marlon Dumas}, \bibinfo{person}{Peter Fettke},
  \bibinfo{person}{Luciano Garc{\'{\i}}a{-}Ba{\~{n}}uelos},
  \bibinfo{person}{Philip Hake}, \bibinfo{person}{Christopher
  Klinkm{\"{u}}ller}, \bibinfo{person}{Henrik Leopold},
  \bibinfo{person}{Andr{\'{e}} Ludwig}, \bibinfo{person}{Peter Loos},
  \bibinfo{person}{Jan Mendling}, \bibinfo{person}{Andreas Oberweis},
  \bibinfo{person}{Andreas Schoknecht}, \bibinfo{person}{Eitam Sheetrit},
  \bibinfo{person}{Tom Thaler}, \bibinfo{person}{Meike Ullrich},
  \bibinfo{person}{Ingo Weber}, {and} \bibinfo{person}{Matthias Weidlich}.}
  \bibinfo{year}{2014}\natexlab{}.
\newblock \showarticletitle{Report: The Process Model Matching Contest 2013}.
  In \bibinfo{booktitle}{{\em Business Process Management Workshops - {BPM}
  2013 International Workshops, Beijing, China, August 26, 2013, Revised
  Papers}} {\em (\bibinfo{series}{Lecture Notes in Business Information
  Processing})}, \bibfield{editor}{\bibinfo{person}{Niels Lohmann},
  \bibinfo{person}{Minseok Song}, {and} \bibinfo{person}{Petia Wohed}} (Eds.),
  Vol.~\bibinfo{volume}{171}. \bibinfo{publisher}{Springer},
  \bibinfo{pages}{442--463}.
\newblock


\bibitem[\protect\citeauthoryear{Chopra, Dalpiaz, Aydemir, Giorgini,
  Mylopoulos, and Singh}{Chopra et~al\mbox{.}}{2014}]%
        {RE-14:Protos}
\bibfield{author}{\bibinfo{person}{Amit~K. Chopra}, \bibinfo{person}{Fabiano
  Dalpiaz}, \bibinfo{person}{F.~Ba{\c{s}}ak Aydemir}, \bibinfo{person}{Paolo
  Giorgini}, \bibinfo{person}{John Mylopoulos}, {and}
  \bibinfo{person}{Munindar~P. Singh}.} \bibinfo{year}{2014}\natexlab{}.
\newblock \showarticletitle{Protos: Foundations for Engineering Innovative
  Sociotechnical Systems}. In \bibinfo{booktitle}{{\em Proceedings of the 18th
  IEEE International Requirements Engineering Conference (RE)}}.
  \bibinfo{publisher}{IEEE Computer Society}, \bibinfo{address}{Karlskrona,
  Sweden}, \bibinfo{pages}{53--62}.
\newblock


\bibitem[\protect\citeauthoryear{Cohn and Hull}{Cohn and Hull}{2009}]%
        {DBLP:journals/debu/CohnH09}
\bibfield{author}{\bibinfo{person}{David Cohn} {and} \bibinfo{person}{Richard
  Hull}.} \bibinfo{year}{2009}\natexlab{}.
\newblock \showarticletitle{Business Artifacts: {A} Data-centric Approach to
  Modeling Business Operations and Processes}.
\newblock \bibinfo{journal}{{\em {IEEE} Data Eng. Bull.\/}}
  \bibinfo{volume}{32}, \bibinfo{number}{3} (\bibinfo{year}{2009}),
  \bibinfo{pages}{3--9}.
\newblock
\showURL{%
\url{http://sites.computer.org/debull/A09sept/david.pdf}}


\bibitem[\protect\citeauthoryear{Decker and Wattenhofer}{Decker and
  Wattenhofer}{2013}]%
        {conf/p2p/DeckerW13}
\bibfield{author}{\bibinfo{person}{Christian Decker} {and}
  \bibinfo{person}{Roger Wattenhofer}.} \bibinfo{year}{2013}\natexlab{}.
\newblock \showarticletitle{Information propagation in the Bitcoin network.}.
  In \bibinfo{booktitle}{{\em P2P}}. \bibinfo{publisher}{IEEE},
  \bibinfo{pages}{1--10}.
\newblock


\bibitem[\protect\citeauthoryear{Decker and Weske}{Decker and Weske}{2011}]%
        {DBLP:journals/is/DeckerW11}
\bibfield{author}{\bibinfo{person}{Gero Decker} {and} \bibinfo{person}{Mathias
  Weske}.} \bibinfo{year}{2011}\natexlab{}.
\newblock \showarticletitle{Interaction-centric modeling of process
  choreographies}.
\newblock \bibinfo{journal}{{\em Inf. Syst.\/}} \bibinfo{volume}{36},
  \bibinfo{number}{2} (\bibinfo{year}{2011}), \bibinfo{pages}{292--312}.
\newblock


\bibitem[\protect\citeauthoryear{Dhillon and Backhouse}{Dhillon and
  Backhouse}{2000}]%
        {dhillon2000technical}
\bibfield{author}{\bibinfo{person}{Gurpreet Dhillon} {and}
  \bibinfo{person}{James Backhouse}.} \bibinfo{year}{2000}\natexlab{}.
\newblock \showarticletitle{Technical opinion: Information system security
  management in the new millennium}.
\newblock \bibinfo{journal}{{\it Commun. ACM}} \bibinfo{volume}{43},
  \bibinfo{number}{7} (\bibinfo{year}{2000}), \bibinfo{pages}{125--128}.
\newblock


\bibitem[\protect\citeauthoryear{Dumas, Rosa, Mendling, and Reijers}{Dumas
  et~al\mbox{.}}{2018}]%
        {DBLP:books/daglib/0031128}
\bibfield{author}{\bibinfo{person}{Marlon Dumas}, \bibinfo{person}{Marcello~La
  Rosa}, \bibinfo{person}{Jan Mendling}, {and} \bibinfo{person}{Hajo~A.
  Reijers}.} \bibinfo{year}{2018}\natexlab{}.
\newblock \bibinfo{booktitle}{{\em Fundamentals of Business Process Management.
  Second Edition}}.
\newblock \bibinfo{publisher}{Springer}.
\newblock


\bibitem[\protect\citeauthoryear{Euzenat and Shvaiko}{Euzenat and
  Shvaiko}{2013}]%
        {DBLP:books/daglib/0032976}
\bibfield{author}{\bibinfo{person}{J{\'{e}}r{\^{o}}me Euzenat} {and}
  \bibinfo{person}{Pavel Shvaiko}.} \bibinfo{year}{2013}\natexlab{}.
\newblock \bibinfo{booktitle}{{\em Ontology Matching, Second Edition}}.
\newblock \bibinfo{publisher}{Springer}.
\newblock
\showISBNx{978-3-642-38720-3}


\bibitem[\protect\citeauthoryear{Eyal and Sirer}{Eyal and Sirer}{2014}]%
        {Eyal2014}
\bibfield{author}{\bibinfo{person}{Ittay Eyal} {and}
  \bibinfo{person}{Emin~G{\"u}n Sirer}.} \bibinfo{year}{2014}\natexlab{}.
\newblock \bibinfo{booktitle}{{\em Majority Is Not Enough: Bitcoin Mining Is
  Vulnerable}}.
\newblock \bibinfo{publisher}{Springer Berlin Heidelberg},
  \bibinfo{address}{Berlin, Heidelberg}, \bibinfo{pages}{436--454}.
\newblock
\showISBNx{978-3-662-45472-5}
\showDOI{%
\url{https://doi.org/10.1007/978-3-662-45472-5_28}}


\bibitem[\protect\citeauthoryear{Fdhila, Indiono, Rinderle{-}Ma, and
  Reichert}{Fdhila et~al\mbox{.}}{2015}]%
        {DBLP:journals/is/FdhilaIRR15}
\bibfield{author}{\bibinfo{person}{Walid Fdhila}, \bibinfo{person}{Conrad
  Indiono}, \bibinfo{person}{Stefanie Rinderle{-}Ma}, {and}
  \bibinfo{person}{Manfred Reichert}.} \bibinfo{year}{2015}\natexlab{}.
\newblock \showarticletitle{Dealing with change in process choreographies:
  Design and implementation of propagation algorithms}.
\newblock \bibinfo{journal}{{\em Inf. Syst.\/}}  \bibinfo{volume}{49}
  (\bibinfo{year}{2015}), \bibinfo{pages}{1--24}.
\newblock
\showDOI{%
\url{https://doi.org/10.1016/j.is.2014.10.004}}


\bibitem[\protect\citeauthoryear{Frantz and Nowostawski}{Frantz and
  Nowostawski}{2016}]%
        {Frantz:ECAS:2016}
\bibfield{author}{\bibinfo{person}{Christopher~K Frantz} {and}
  \bibinfo{person}{Mariusz Nowostawski}.} \bibinfo{year}{2016}\natexlab{}.
\newblock \showarticletitle{From Institutions to Code: Towards Automated
  Generation of Smart Contracts}. In \bibinfo{booktitle}{{\em Workshop on
  Engineering Collective Adaptive Systems (eCAS), co-located with SASO,
  Augsburg}}.
\newblock


\bibitem[\protect\citeauthoryear{Gal}{Gal}{2011}]%
        {gal2011uncertain}
\bibfield{author}{\bibinfo{person}{Avigdor Gal}.}
  \bibinfo{year}{2011}\natexlab{}.
\newblock \bibinfo{booktitle}{{\em Uncertain schema matching}}.
\newblock \bibinfo{publisher}{Morgan \& Claypool Publishers}.
\newblock


\bibitem[\protect\citeauthoryear{Garc{\'i}a{-}Ba{\~{n}}uelos, Ponomarev, Dumas,
  and Weber}{Garc{\'i}a{-}Ba{\~{n}}uelos et~al\mbox{.}}{2017}]%
        {Garcia:2017:BPM}
\bibfield{author}{\bibinfo{person}{Luciano Garc{\'i}a{-}Ba{\~{n}}uelos},
  \bibinfo{person}{Alexander Ponomarev}, \bibinfo{person}{Marlon Dumas}, {and}
  \bibinfo{person}{Ingo Weber}.} \bibinfo{year}{2017}\natexlab{}.
\newblock \showarticletitle{Optimized Execution of Business Processes on
  Blockchain}. In \bibinfo{booktitle}{{\em BPM'17: International Conference on
  Business Process Management}}. \bibinfo{address}{Barcelona, Spain}.
\newblock


\bibitem[\protect\citeauthoryear{Gervais, Karame, W\"{u}st, Glykantzis,
  Ritzdorf, and Capkun}{Gervais et~al\mbox{.}}{2016}]%
        {Gervais:2016:SPP:2976749.2978341}
\bibfield{author}{\bibinfo{person}{Arthur Gervais}, \bibinfo{person}{Ghassan~O.
  Karame}, \bibinfo{person}{Karl W\"{u}st}, \bibinfo{person}{Vasileios
  Glykantzis}, \bibinfo{person}{Hubert Ritzdorf}, {and} \bibinfo{person}{Srdjan
  Capkun}.} \bibinfo{year}{2016}\natexlab{}.
\newblock \showarticletitle{On the Security and Performance of Proof of Work
  Blockchains}. In \bibinfo{booktitle}{{\em Proceedings of the 2016 ACM SIGSAC
  Conference on Computer and Communications Security}} {\em
  (\bibinfo{series}{CCS '16})}. \bibinfo{publisher}{ACM}, \bibinfo{address}{New
  York, NY, USA}, \bibinfo{pages}{3--16}.
\newblock
\showISBNx{978-1-4503-4139-4}
\showDOI{%
\url{https://doi.org/10.1145/2976749.2978341}}


\bibitem[\protect\citeauthoryear{Guo and Liang}{Guo and Liang}{2016}]%
        {guo2016blockchain}
\bibfield{author}{\bibinfo{person}{Ye Guo} {and} \bibinfo{person}{Chen Liang}.}
  \bibinfo{year}{2016}\natexlab{}.
\newblock \showarticletitle{Blockchain application and outlook in the banking
  industry}.
\newblock \bibinfo{journal}{{\em Financial Innovation\/}} \bibinfo{volume}{2},
  \bibinfo{number}{1} (\bibinfo{year}{2016}), \bibinfo{pages}{24}.
\newblock


\bibitem[\protect\citeauthoryear{Hull, Batra, Chen, Deutsch, III, and
  Vianu}{Hull et~al\mbox{.}}{2016}]%
        {DBLP:conf/icsoc/HullBCDHV16}
\bibfield{author}{\bibinfo{person}{Richard Hull}, \bibinfo{person}{Vishal~S.
  Batra}, \bibinfo{person}{Yi{-}Min Chen}, \bibinfo{person}{Alin Deutsch},
  \bibinfo{person}{Fenno F. Terry~Heath III}, {and} \bibinfo{person}{Victor
  Vianu}.} \bibinfo{year}{2016}\natexlab{}.
\newblock \showarticletitle{Towards a Shared Ledger Business Collaboration
  Language Based on Data-Aware Processes}. In \bibinfo{booktitle}{{\em
  Service-Oriented Computing - 14th International Conference, {ICSOC} 2016,
  Banff, AB, Canada, October 10-13, 2016, Proceedings}} {\em
  (\bibinfo{series}{Lecture Notes in Computer Science})},
  \bibfield{editor}{\bibinfo{person}{Quan~Z. Sheng}, \bibinfo{person}{Eleni
  Stroulia}, \bibinfo{person}{Samir Tata}, {and} \bibinfo{person}{Sami Bhiri}}
  (Eds.), Vol.~\bibinfo{volume}{9936}. \bibinfo{publisher}{Springer},
  \bibinfo{pages}{18--36}.
\newblock


\bibitem[\protect\citeauthoryear{Hurlburt}{Hurlburt}{2016}]%
        {hurlburt2016might}
\bibfield{author}{\bibinfo{person}{George Hurlburt}.}
  \bibinfo{year}{2016}\natexlab{}.
\newblock \showarticletitle{Might the Blockchain Outlive Bitcoin?}
\newblock \bibinfo{journal}{{\em IT Professional\/}} \bibinfo{volume}{18},
  \bibinfo{number}{2} (\bibinfo{year}{2016}), \bibinfo{pages}{12--16}.
\newblock


\bibitem[\protect\citeauthoryear{Kosba, Miller, Shi, Wen, and
  Papamanthou}{Kosba et~al\mbox{.}}{2016}]%
        {7546538}
\bibfield{author}{\bibinfo{person}{A. Kosba}, \bibinfo{person}{A. Miller},
  \bibinfo{person}{E. Shi}, \bibinfo{person}{Z. Wen}, {and} \bibinfo{person}{C.
  Papamanthou}.} \bibinfo{year}{2016}\natexlab{}.
\newblock \showarticletitle{Hawk: The Blockchain Model of Cryptography and
  Privacy-Preserving Smart Contracts}. In \bibinfo{booktitle}{{\em 2016 IEEE
  Symposium on Security and Privacy (SP)}}. \bibinfo{pages}{839--858}.
\newblock
\showDOI{%
\url{https://doi.org/10.1109/SP.2016.55}}


\bibitem[\protect\citeauthoryear{L{\'{o}}pez{-}Pintado,
  Garc{\'{\i}}a{-}Ba{\~{n}}uelos, Dumas, and Weber}{L{\'{o}}pez{-}Pintado
  et~al\mbox{.}}{2017}]%
        {DBLP:conf/bpm/Lopez-PintadoGD17}
\bibfield{author}{\bibinfo{person}{Orlenys L{\'{o}}pez{-}Pintado},
  \bibinfo{person}{Luciano Garc{\'{\i}}a{-}Ba{\~{n}}uelos},
  \bibinfo{person}{Marlon Dumas}, {and} \bibinfo{person}{Ingo Weber}.}
  \bibinfo{year}{2017}\natexlab{}.
\newblock \showarticletitle{Caterpillar: {A} Blockchain-Based Business Process
  Management System}. In \bibinfo{booktitle}{{\em Proceedings of the {BPM} Demo
  Track and {BPM} Dissertation Award co-located with 15th International
  Conference on Business Process Modeling {(BPM} 2017), Barcelona, Spain,
  September 13, 2017.}} {\em (\bibinfo{series}{{CEUR} Workshop Proceedings})},
  \bibfield{editor}{\bibinfo{person}{Robert Claris{\'{o}}},
  \bibinfo{person}{Henrik Leopold}, \bibinfo{person}{Jan Mendling},
  \bibinfo{person}{Wil M.~P. van~der Aalst}, \bibinfo{person}{Akhil Kumar},
  \bibinfo{person}{Brian~T. Pentland}, {and} \bibinfo{person}{Mathias Weske}}
  (Eds.), Vol.~\bibinfo{volume}{1920}. \bibinfo{publisher}{CEUR-WS.org}.
\newblock
\showURL{%
\url{http://ceur-ws.org/Vol-1920}}


\bibitem[\protect\citeauthoryear{Marin, Hull, and Vacul{\'{\i}}n}{Marin
  et~al\mbox{.}}{2012}]%
        {DBLP:conf/bpm/MarinHV12}
\bibfield{author}{\bibinfo{person}{Mike Marin}, \bibinfo{person}{Richard Hull},
  {and} \bibinfo{person}{Roman Vacul{\'{\i}}n}.}
  \bibinfo{year}{2012}\natexlab{}.
\newblock \showarticletitle{Data Centric {BPM} and the Emerging Case Management
  Standard: {A} Short Survey}. In \bibinfo{booktitle}{{\em Business Process
  Management Workshops, Tallinn, Estonia, September 3, 2012. Revised Papers}}.
  \bibinfo{publisher}{Springer}, \bibinfo{pages}{24--30}.
\newblock


\bibitem[\protect\citeauthoryear{Mendling and Hafner}{Mendling and
  Hafner}{2008}]%
        {mendling2008ws}
\bibfield{author}{\bibinfo{person}{Jan Mendling} {and} \bibinfo{person}{Michael
  Hafner}.} \bibinfo{year}{2008}\natexlab{}.
\newblock \showarticletitle{From {WS-CDL} choreography to {BPEL} process
  orchestration}.
\newblock \bibinfo{journal}{{\em J. Enterprise Information Management\/}}
  \bibinfo{volume}{21}, \bibinfo{number}{5} (\bibinfo{year}{2008}),
  \bibinfo{pages}{525--542}.
\newblock


\bibitem[\protect\citeauthoryear{Mougayar}{Mougayar}{2016}]%
        {mougayar2016business}
\bibfield{author}{\bibinfo{person}{W Mougayar}.}
  \bibinfo{year}{2016}\natexlab{}.
\newblock \bibinfo{booktitle}{{\em The Business Blockchain: Promise, Practice,
  and Application of the Next Internet Technology}}.
\newblock \bibinfo{publisher}{Wiley}.
\newblock


\bibitem[\protect\citeauthoryear{Nakamoto}{Nakamoto}{2008}]%
        {nakamoto2008bitcoin}
\bibfield{author}{\bibinfo{person}{Satoshi Nakamoto}.}
  \bibinfo{year}{2008}\natexlab{}.
\newblock \bibinfo{title}{Bitcoin: A peer-to-peer electronic cash system}.
\newblock   (\bibinfo{year}{2008}).
\newblock


\bibitem[\protect\citeauthoryear{Natoli and Gramoli}{Natoli and
  Gramoli}{2017}]%
        {NG:2017:DSN}
\bibfield{author}{\bibinfo{person}{Christopher Natoli} {and}
  \bibinfo{person}{Vincent Gramoli}.} \bibinfo{year}{2017}\natexlab{}.
\newblock \showarticletitle{The Balance Attack or Why Forkable Blockchains Are
  Ill-Suited for Consortium}. In \bibinfo{booktitle}{{\em The 47th IEEE/IFIP
  International Conference on Dependable Systems and Networks (DSN'17)}}.
  \bibinfo{publisher}{IEEE}.
\newblock


\bibitem[\protect\citeauthoryear{Nofer, Gomber, Hinz, and Schiereck}{Nofer
  et~al\mbox{.}}{2017}]%
        {DBLP:journals/bise/NoferGHS17}
\bibfield{author}{\bibinfo{person}{Michael Nofer}, \bibinfo{person}{Peter
  Gomber}, \bibinfo{person}{Oliver Hinz}, {and} \bibinfo{person}{Dirk
  Schiereck}.} \bibinfo{year}{2017}\natexlab{}.
\newblock \showarticletitle{Blockchain}.
\newblock \bibinfo{journal}{{\em Business {\&} Information Systems
  Engineering\/}} \bibinfo{volume}{59}, \bibinfo{number}{3}
  (\bibinfo{year}{2017}), \bibinfo{pages}{183--187}.
\newblock
\showDOI{%
\url{https://doi.org/10.1007/s12599-017-0467-3}}


\bibitem[\protect\citeauthoryear{Norta}{Norta}{2015}]%
        {Norta:2015:BIR}
\bibfield{author}{\bibinfo{person}{Alex Norta}.}
  \bibinfo{year}{2015}\natexlab{}.
\newblock \showarticletitle{Creation of Smart-Contracting Collaborations for
  Decentralized Autonomous Organizations}. In \bibinfo{booktitle}{{\em BIR'15:
  International Conference on Perspectives in Business Informatics Research}}.
  \bibinfo{pages}{3--17}.
\newblock


\bibitem[\protect\citeauthoryear{Norta}{Norta}{2016}]%
        {Norta:2016:ICACDS}
\bibfield{author}{\bibinfo{person}{Alex Norta}.}
  \bibinfo{year}{2016}\natexlab{}.
\newblock \showarticletitle{Designing a Smart-Contract Application Layer for
  Transacting Decentralized Autonomous Organizations}. In
  \bibinfo{booktitle}{{\em ICACDS'16: International Conference on Advances in
  Computing and Data Sciences}}.
\newblock


\bibitem[\protect\citeauthoryear{Prybila, Schulte, Hochreiner, and
  Weber}{Prybila et~al\mbox{.}}{2017}]%
        {PSHW17}
\bibfield{author}{\bibinfo{person}{Christoph Prybila}, \bibinfo{person}{Stefan
  Schulte}, \bibinfo{person}{Christoph Hochreiner}, {and} \bibinfo{person}{Ingo
  Weber}.} \bibinfo{year}{2017}\natexlab{}.
\newblock \bibinfo{booktitle}{{\em {Runtime Verification for Business Processes
  Utilizing the Bitcoin Blockchain}}}.
\newblock \bibinfo{type}{arXiv report} 1706.04404.
  \bibinfo{institution}{arXiv}.
\newblock
\showURL{%
\url{https://arxiv.org/abs/1706.04404}}


\bibitem[\protect\citeauthoryear{Ramberg}{Ramberg}{2011}]%
        {ramberg2011icc}
\bibfield{author}{\bibinfo{person}{Jan Ramberg}.}
  \bibinfo{year}{2011}\natexlab{}.
\newblock \showarticletitle{{ICC Guide to Incoterms 2010}}. ICC.
\newblock


\bibitem[\protect\citeauthoryear{Reichert and Weber}{Reichert and
  Weber}{2012}]%
        {DBLP:books/daglib/0030179}
\bibfield{author}{\bibinfo{person}{Manfred Reichert} {and}
  \bibinfo{person}{Barbara Weber}.} \bibinfo{year}{2012}\natexlab{}.
\newblock \bibinfo{booktitle}{{\em Enabling Flexibility in Process-Aware
  Information Systems - Challenges, Methods, Technologies}}.
\newblock \bibinfo{publisher}{Springer}.
\newblock
\showISBNx{978-3-642-30408-8}
\showDOI{%
\url{https://doi.org/10.1007/978-3-642-30409-5}}


\bibitem[\protect\citeauthoryear{Rosemann and vom Brocke}{Rosemann and vom
  Brocke}{2015}]%
        {rosemann2015six}
\bibfield{author}{\bibinfo{person}{Michael Rosemann} {and} \bibinfo{person}{Jan
  vom Brocke}.} \bibinfo{year}{2015}\natexlab{}.
\newblock \showarticletitle{The six core elements of business process
  management}.
\newblock In \bibinfo{booktitle}{{\em Handbook on Business Process Management
  1}}. \bibinfo{publisher}{Springer}, \bibinfo{pages}{105--122}.
\newblock


\bibitem[\protect\citeauthoryear{Shleifer and Vishny}{Shleifer and
  Vishny}{1997}]%
        {shleifer1997survey}
\bibfield{author}{\bibinfo{person}{Andrei Shleifer} {and}
  \bibinfo{person}{Robert~W Vishny}.} \bibinfo{year}{1997}\natexlab{}.
\newblock \showarticletitle{A survey of corporate governance}.
\newblock \bibinfo{journal}{{\em The journal of finance\/}}
  \bibinfo{volume}{52}, \bibinfo{number}{2} (\bibinfo{year}{1997}),
  \bibinfo{pages}{737--783}.
\newblock


\bibitem[\protect\citeauthoryear{Staples, Chen, Falamaki, Ponomarev, Rimba,
  Tran, Weber, Xu, and Zhu}{Staples et~al\mbox{.}}{2017}]%
        {Risks-Blockchain-2017}
\bibfield{author}{\bibinfo{person}{M. Staples}, \bibinfo{person}{S. Chen},
  \bibinfo{person}{S. Falamaki}, \bibinfo{person}{A. Ponomarev},
  \bibinfo{person}{P. Rimba}, \bibinfo{person}{A.~B. Tran}, \bibinfo{person}{I.
  Weber}, \bibinfo{person}{X. Xu}, {and} \bibinfo{person}{L. Zhu}.}
  \bibinfo{year}{2017}\natexlab{}.
\newblock \bibinfo{booktitle}{{\em Risks and opportunities for systems using
  blockchain and smart contracts}}.
\newblock \bibinfo{type}{{T}echnical {R}eport}. \bibinfo{institution}{Data61
  (CSIRO), Sydney}.
\newblock


\bibitem[\protect\citeauthoryear{{\v{S}}temberger, Buh, Glavan, and
  Mendling}{{\v{S}}temberger et~al\mbox{.}}{2017}]%
        {stemberger-2017}
\bibfield{author}{\bibinfo{person}{Mojca~Indihar {\v{S}}temberger},
  \bibinfo{person}{Brina Buh}, \bibinfo{person}{Ljubica~Milanovic Glavan},
  {and} \bibinfo{person}{Jan Mendling}.} \bibinfo{year}{2017}\natexlab{}.
\newblock \showarticletitle{Propositions on the Interaction of Organizational
  Culture with other Factors in the context of BPM Adoption}.
\newblock \bibinfo{journal}{{\em Business Process Management Journal\/}}
  \bibinfo{volume}{23} (\bibinfo{year}{2017}).
\newblock


\bibitem[\protect\citeauthoryear{Subramanian, Thiran, Narendra,
  Most{\'{e}}faoui, and Maamar}{Subramanian et~al\mbox{.}}{2008}]%
        {DBLP:conf/saint/SattanathanTNMM08}
\bibfield{author}{\bibinfo{person}{Sattanathan Subramanian},
  \bibinfo{person}{Philippe Thiran}, \bibinfo{person}{Nanjangud Narendra},
  \bibinfo{person}{Ghita Most{\'{e}}faoui}, {and} \bibinfo{person}{Zakaria
  Maamar}.} \bibinfo{year}{2008}\natexlab{}.
\newblock \showarticletitle{On the Enhancement of {BPEL} Engines for
  Self-Healing Composite Web Services}. In \bibinfo{booktitle}{{\em Proc. SAINT
  Symposium}}. \bibinfo{pages}{33--39}.
\newblock


\bibitem[\protect\citeauthoryear{Swan}{Swan}{2015}]%
        {swan2015blockchain}
\bibfield{author}{\bibinfo{person}{Melanie Swan}.}
  \bibinfo{year}{2015}\natexlab{}.
\newblock \bibinfo{booktitle}{{\em Blockchain: Blueprint for a new economy}}.
\newblock \bibinfo{publisher}{O'Reilly Media, Inc.}
\newblock


\bibitem[\protect\citeauthoryear{Szabo}{Szabo}{1997}]%
        {szabo1997formalizing}
\bibfield{author}{\bibinfo{person}{Nick Szabo}.}
  \bibinfo{year}{1997}\natexlab{}.
\newblock \showarticletitle{Formalizing and securing relationships on public
  networks}.
\newblock \bibinfo{journal}{{\em First Monday\/}} \bibinfo{volume}{2},
  \bibinfo{number}{9} (\bibinfo{year}{1997}).
\newblock


\bibitem[\protect\citeauthoryear{Tassey}{Tassey}{2000}]%
        {tassey2000standardization}
\bibfield{author}{\bibinfo{person}{Gregory Tassey}.}
  \bibinfo{year}{2000}\natexlab{}.
\newblock \showarticletitle{Standardization in technology-based markets}.
\newblock \bibinfo{journal}{{\em Research policy\/}} \bibinfo{volume}{29},
  \bibinfo{number}{4} (\bibinfo{year}{2000}), \bibinfo{pages}{587--602}.
\newblock


\bibitem[\protect\citeauthoryear{Telang and Singh}{Telang and Singh}{2012}]%
        {AAMAS-Comma-12}
\bibfield{author}{\bibinfo{person}{Pankaj~R. Telang} {and}
  \bibinfo{person}{Munindar~P. Singh}.} \bibinfo{year}{2012}\natexlab{}.
\newblock \showarticletitle{Comma: A Commitment-Based Business Modeling
  Methodology and its Empirical Evaluation}. \bibinfo{publisher}{IFAAMAS},
  \bibinfo{address}{Valencia, Spain}, \bibinfo{pages}{1073--1080}.
\newblock


\bibitem[\protect\citeauthoryear{Tran, Xu, Weber, Staples, and Rimba}{Tran
  et~al\mbox{.}}{2017}]%
        {Tran:2017:CAISE}
\bibfield{author}{\bibinfo{person}{An~Binh Tran}, \bibinfo{person}{Xiwei Xu},
  \bibinfo{person}{Ingo Weber}, \bibinfo{person}{Mark Staples}, {and}
  \bibinfo{person}{Paul Rimba}.} \bibinfo{year}{2017}\natexlab{}.
\newblock \showarticletitle{Regerator: a Registry Generator for Blockchain}. In
  \bibinfo{booktitle}{{\em CAiSE'17: International Conference on Advanced
  Information Systems Engineering, Forum Track (demo)}}.
\newblock


\bibitem[\protect\citeauthoryear{van~der Aalst}{van~der Aalst}{2016}]%
        {DBLP:books/sp/Aalst16}
\bibfield{author}{\bibinfo{person}{Wil M.~P. van~der Aalst}.}
  \bibinfo{year}{2016}\natexlab{}.
\newblock \bibinfo{booktitle}{{\em Process Mining - Data Science in Action,
  Second Edition}}.
\newblock \bibinfo{publisher}{Springer}.
\newblock
\showISBNx{978-3-662-49850-7}


\bibitem[\protect\citeauthoryear{van~der Aalst, Dumas, Ouyang, Rozinat, and
  Verbeek}{van~der Aalst et~al\mbox{.}}{2008}]%
        {DBLP:journals/toit/AalstDORV08}
\bibfield{author}{\bibinfo{person}{Wil M.~P. van~der Aalst},
  \bibinfo{person}{Marlon Dumas}, \bibinfo{person}{Chun Ouyang},
  \bibinfo{person}{Anne Rozinat}, {and} \bibinfo{person}{Eric Verbeek}.}
  \bibinfo{year}{2008}\natexlab{}.
\newblock \showarticletitle{Conformance checking of service behavior}.
\newblock \bibinfo{journal}{{\em {ACM} Trans. Internet Techn.\/}}
  \bibinfo{volume}{8}, \bibinfo{number}{3} (\bibinfo{year}{2008}).
\newblock


\bibitem[\protect\citeauthoryear{van~der Aalst and Weske}{van~der Aalst and
  Weske}{2001}]%
        {DBLP:conf/caise/AalstW01}
\bibfield{author}{\bibinfo{person}{Wil M.~P. van~der Aalst} {and}
  \bibinfo{person}{Mathias Weske}.} \bibinfo{year}{2001}\natexlab{}.
\newblock \showarticletitle{The {P2P} Approach to Interorganizational
  Workflows}. In \bibinfo{booktitle}{{\em Proc. CAiSE}}.
  \bibinfo{pages}{140--156}.
\newblock


\bibitem[\protect\citeauthoryear{Vanwersch, Shahzad, Vanderfeesten, Vanhaecht,
  Grefen, Pintelon, Mendling, van Merode, and Reijers}{Vanwersch
  et~al\mbox{.}}{2016}]%
        {DBLP:journals/bise/VanwerschSVVGPM16}
\bibfield{author}{\bibinfo{person}{Rob J.~B. Vanwersch},
  \bibinfo{person}{Khurram Shahzad}, \bibinfo{person}{Irene T.~P.
  Vanderfeesten}, \bibinfo{person}{Kris Vanhaecht}, \bibinfo{person}{Paul W.
  P.~J. Grefen}, \bibinfo{person}{Liliane Pintelon}, \bibinfo{person}{Jan
  Mendling}, \bibinfo{person}{Godefridus~G. van Merode}, {and}
  \bibinfo{person}{Hajo~A. Reijers}.} \bibinfo{year}{2016}\natexlab{}.
\newblock \showarticletitle{A Critical Evaluation and Framework of Business
  Process Improvement Methods}.
\newblock \bibinfo{journal}{{\em Business {\&} Information Systems
  Engineering\/}} \bibinfo{volume}{58}, \bibinfo{number}{1}
  (\bibinfo{year}{2016}), \bibinfo{pages}{43--53}.
\newblock
\showDOI{%
\url{https://doi.org/10.1007/s12599-015-0417-x}}


\bibitem[\protect\citeauthoryear{Venkatesh, Morris, Davis, and Davis}{Venkatesh
  et~al\mbox{.}}{2003}]%
        {venkatesh2003user}
\bibfield{author}{\bibinfo{person}{Viswanath Venkatesh},
  \bibinfo{person}{Michael~G Morris}, \bibinfo{person}{Gordon~B Davis}, {and}
  \bibinfo{person}{Fred~D Davis}.} \bibinfo{year}{2003}\natexlab{}.
\newblock \showarticletitle{User acceptance of information technology: Toward a
  unified view}.
\newblock \bibinfo{journal}{{\em MIS quarterly\/}} (\bibinfo{year}{2003}),
  \bibinfo{pages}{425--478}.
\newblock


\bibitem[\protect\citeauthoryear{vom Brocke and Sinnl}{vom Brocke and
  Sinnl}{2011}]%
        {brocke2011culture}
\bibfield{author}{\bibinfo{person}{Jan vom Brocke} {and}
  \bibinfo{person}{Theresa Sinnl}.} \bibinfo{year}{2011}\natexlab{}.
\newblock \showarticletitle{Culture in business process management: a
  literature review}.
\newblock \bibinfo{journal}{{\em Business Process Management Journal\/}}
  \bibinfo{volume}{17}, \bibinfo{number}{2} (\bibinfo{year}{2011}),
  \bibinfo{pages}{357--378}.
\newblock


\bibitem[\protect\citeauthoryear{Walport}{Walport}{2016}]%
        {walport2016distributed}
\bibfield{author}{\bibinfo{person}{M Walport}.}
  \bibinfo{year}{2016}\natexlab{}.
\newblock \showarticletitle{Distributed Ledger Technology: Beyond Blockchain}.
\newblock \bibinfo{journal}{{\em UK Government Office for Science, Tech.
  Rep\/}}  \bibinfo{volume}{19, 2016} (\bibinfo{year}{2016}).
\newblock


\bibitem[\protect\citeauthoryear{Weber, Gramoli, Staples, Ponomarev, Holz,
  Tran, and Rimba}{Weber et~al\mbox{.}}{2017}]%
        {Weber:2017:SRDS}
\bibfield{author}{\bibinfo{person}{Ingo Weber}, \bibinfo{person}{Vincent
  Gramoli}, \bibinfo{person}{Mark Staples}, \bibinfo{person}{Alex Ponomarev},
  \bibinfo{person}{Ralph Holz}, \bibinfo{person}{{An Binh} Tran}, {and}
  \bibinfo{person}{Paul Rimba}.} \bibinfo{year}{2017}\natexlab{}.
\newblock \showarticletitle{On Availability for Blockchain-Based Systems}. In
  \bibinfo{booktitle}{{\em SRDS'17: IEEE International Symposium on Reliable
  Distributed Systems}}.
\newblock


\bibitem[\protect\citeauthoryear{Weber, Haller, and M\"ulle}{Weber
  et~al\mbox{.}}{2008}]%
        {2008-Weber-IJBPIM}
\bibfield{author}{\bibinfo{person}{Ingo Weber}, \bibinfo{person}{Jochen
  Haller}, {and} \bibinfo{person}{Jutta M\"ulle}.}
  \bibinfo{year}{2008}\natexlab{}.
\newblock \showarticletitle{Automated Derivation of Executable Business
  Processes from Choreograpies in Virtual Organizations}.
\newblock \bibinfo{journal}{{\em International Journal of Business Process
  Integration and Management (IJBPIM)\/}} \bibinfo{volume}{3},
  \bibinfo{number}{2} (\bibinfo{year}{2008}), \bibinfo{pages}{85--95}.
\newblock


\bibitem[\protect\citeauthoryear{Weber, Xu, Riveret, Governatori, Ponomarev,
  and Mendling}{Weber et~al\mbox{.}}{2016}]%
        {Weber:2016:BPM}
\bibfield{author}{\bibinfo{person}{Ingo Weber}, \bibinfo{person}{Xiwei Xu},
  \bibinfo{person}{Regis Riveret}, \bibinfo{person}{Guido Governatori},
  \bibinfo{person}{Alexander Ponomarev}, {and} \bibinfo{person}{Jan Mendling}.}
  \bibinfo{year}{2016}\natexlab{}.
\newblock \showarticletitle{Untrusted Business Process Monitoring and Execution
  Using Blockchain}. In \bibinfo{booktitle}{{\em Business Process Management -
  14th International Conference, {BPM} 2016, Rio de Janeiro, Brazil, September
  18-22, 2016. Proceedings}} {\em (\bibinfo{series}{Lecture Notes in Computer
  Science})}, Vol.~\bibinfo{volume}{9850}. \bibinfo{publisher}{Springer},
  \bibinfo{pages}{329--347}.
\newblock


\bibitem[\protect\citeauthoryear{Yli-Huumo, Ko, Choi, Park, and
  Smolander}{Yli-Huumo et~al\mbox{.}}{2016}]%
        {Yli-Huumo:2016:PlosOne}
\bibfield{author}{\bibinfo{person}{J Yli-Huumo}, \bibinfo{person}{D Ko},
  \bibinfo{person}{S Choi}, \bibinfo{person}{S Park}, {and} \bibinfo{person}{K
  Smolander}.} \bibinfo{year}{2016}\natexlab{}.
\newblock \showarticletitle{Where Is Current Research on Blockchain Technology?
  -- {A} Systematic Review}.
\newblock \bibinfo{journal}{{\em PLoSONE\/}} \bibinfo{volume}{11},
  \bibinfo{number}{10} (\bibinfo{year}{2016}), \bibinfo{pages}{e0163477}.
\newblock


\bibitem[\protect\citeauthoryear{Zeng, Benatallah, Ngu, Dumas, Kalagnanam, and
  Chang}{Zeng et~al\mbox{.}}{2004}]%
        {DBLP:journals/tse/ZengBNDKC04}
\bibfield{author}{\bibinfo{person}{Liangzhao Zeng}, \bibinfo{person}{Boualem
  Benatallah}, \bibinfo{person}{Anne Ngu}, \bibinfo{person}{Marlon Dumas},
  \bibinfo{person}{Jayant Kalagnanam}, {and} \bibinfo{person}{Henry Chang}.}
  \bibinfo{year}{2004}\natexlab{}.
\newblock \showarticletitle{QoS-Aware Middleware for Web Services Composition}.
\newblock \bibinfo{journal}{{\em {IEEE} TSE\/}} \bibinfo{volume}{30},
  \bibinfo{number}{5} (\bibinfo{year}{2004}), \bibinfo{pages}{311--327}.
\newblock


\end{thebibliography}
